# Large Area Vapor Phase Growth and Characterization of MoS$_2$ Atomic Layers on SiO$_2$ Substrate

Yongjie Zhan[1#], Zheng Liu[1#], Sina Najmaei[1], Pulickel M. Ajayan[1]* & Jun Lou[1]*

1. Department of Mechanical Engineering & Materials Science, Rice University, Houston, Texas 77005, US

[#] These authors contributed equally to this work

*Corresponding authors:

Email: ajayan@rice.edu (Pulickel M. Ajayan); jlou@rice.edu (Jun Lou).

**Table of Contents Graphic**

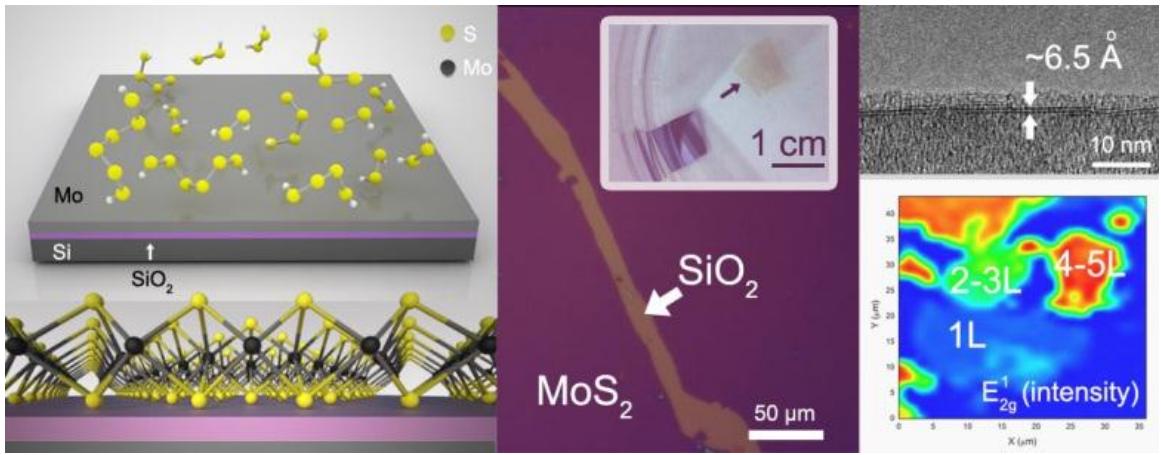

**Abstract**

Monolayer Molybdenum disulfide ($MoS_2$), a two-dimensional crystal with a direct bandgap, is a promising candidate for 2D nanoelectronic devices complementing graphene. There have been recent attempts to produce $MoS_2$ layers via chemical and mechanical exfoliation of bulk material. Here we demonstrate the large area growth of $MoS_2$ atomic layers on $SiO_2$ substrates by a scalable chemical vapor deposition (CVD) method. The as-prepared samples can either be readily utilized for further device fabrication or be easily released from $SiO_2$ and transferred to arbitrary substrates. High resolution transmission electron microscopy and Raman spectroscopy on the as grown films of $MoS_2$ indicate that the number of layers range from single layer to a few layers. Our results on the direct growth of $MoS_2$ layers on dielectric leading to facile device fabrication possibilities show the expanding set of useful 2D atomic layers, on the heels of graphene, which can be controllably synthesized and manipulated for many applications.

Inspired by recent success in graphene based research[1-3], monolayer or few-layer nanostructures derived from other layered materials such as hexagonal Boron Nitride (h-BN) and transition-metal dichalcogenides including $MoS_2$, $WS_2$ etc. have received increasing attention due to their potential for a range of applications[4-7]. Unlike conductive graphene and insulating h-BN, atomic layered $MoS_2$ is a semiconductor material with a direct bandgap, offering possibilities of fabricating high performance devices with low power consumption in a more straight-forward manner[8]. In a recent effort to fabricate single-layer $MoS_2$ transistors, impressive mobility of at least 200 $cm^2V^{-1} s^{-1}$ has been demonstrated using a halfnium oxide ($HfO_2$) gate dielectric[8], a big increase from 0.5 – 3 $cm^2V^{-1} s^{-1}$ reported earlier using a silicon oxide gate dielectric[1]. However, the traditional mechanical exfoliation method is still employed to obtain the $MoS_2$ atomic layer with rather modest foot-print, limiting its usefulness in a commercially viable device. Liquid exfoliation of layered materials including $MoS_2$ has been proposed to be a promising large scale synthesis method for two-dimensional nanosheet[9]. Although it is quite facile to create hybrid dispersions or composites using this method, its application into device applications still needs further development. Other methods including hydrothermal methods that were employed to synthesize $MoS_2$ nanosheet have similar limitations[10-12]. Therefore, large area synthesis of monolayer and few-layer $MoS_2$ that is compatible with current micro- or nano-fabrication processes will greatly facilitate the integration of this fascinating material into future device applications. In the present work, we report that a rather simple and direct elemental reaction between Mo and S can produce large area good quality $MoS_2$ atomic layers on $SiO_2$ substrates.

In a typical procedure, samples (Mo thin films deposited on $SiO_2$ substrates) placed in a ceramic boat were placed in the center of a tube furnace (Lindberg, Blue M, quartz tube). Another ceramic boat holding pure sulfur (1-2g, Fisher Scientific, USP grade) was placed in the upwind low temperature zone in the quartz tube. During the reaction, the temperature in the low temperature zone were controlled to be a little above the melting point of sulfur (113°C). The quartz tube was first kept in a flowing protective atmosphere of high purity $N_2$, the flow rate of which was set at 150-200 sccm. After 15 minutes of $N_2$ purging, the furnace temperature was gradually increased from room temperature to 500 °C in 30 minutes. Then the temperature was increased again from 500 °C to 750 °C in 90 minutes and was kept at 750 °C for 10 minutes before cooled down to room temperature in 120 minutes. Figure s1 shows a schematic illustration of the reaction condition of this CVD process. Raman spectroscopy (Renishaw inVia) was performed with 514.5 nm laser excitation. Scanning electron microscope (FEI Quanta 400) and high resolution transmission electron microscopy (HRTEM, JEOL-2100) equipped with electron energy loss spectrum (EELS) and GIF filter were employed for imaging and chemical analysis of the samples. X-ray photoelecton spectroscopy (XPS, PHI Quantera) was performed using monochromatic aluminum KR X-rays. MultiPak software was used for XPS data analyses.

As illustrated in Fig. 1a, thin layer of Mo (typical thickness 1 ~ 5 nm) was pre-deposited on $SiO_2$/Si by e-beam evaporation at a rate of ~0.1A/s. Sulfur was introduced and reacted with Mo at 750°C forming very thin $MoS_2$ film (form single layer to few layers), as illustrated in Fig. 1b. The as-prepared $MoS_2$ atomic layers on $SiO_2$ substrates are readily available for further characterizations as well as device fabrications. It is also easy to transfer the thin layers onto arbitrary substrates by etching away the $SiO_2$ using KOH solution (~15M). Fig. 1c shows a released $MoS_2$ atomic layers floating on the surface of the alkaline solution. The lateral size of the $MoS_2$ layers is simply dependent on the size of the substrates used (~0.8cm×0.8cm as shown in Fig. 1c), suggesting that the process is scalable and films of any size can be grown with good uniformity. The thickness of the $MoS_2$ atomic layer grown directly relates to the thickness of the pre-deposited Mo metal on the substrate and hence the thickness of the layers can be controlled. The $MoS_2$ atomic layers can then be transferred onto arbitrary substrates (including TEM sample grids) for further characterizations and processing. Figure 1d shows an optical image captured from the edge of a typical $MoS_2$ on a $SiO_2$ substrate (285nm). The light purple area in the top-right corner marked by a yellow arrow shows a very thin area (1-2 layers), while most of other areas are few-layered $MoS_2$ in purple. Fig. 1e shows the corresponding SEM image. The morphologies reveal that the on-site growth of $MoS_2$ on $SiO_2$ substrate can produce very thin, continuous and uniform atomic layers. Fig. 1f shows a SEM image of a large size $MoS_2$ in uniform. More optical and SEM images can be found in supporting information Figs. S1. In our experiments, we tried various substrates (Si, $SiO_2$, $Al_2O_3$, Cr, Au, Au / Cr bi-layer) to deposit Mo on them using e-beam evaporation. All other substrates ($Al_2O_3$, Cr, Au and Au/Cr) were pre-deposited thin films on silicon wafers. The growth of $MoS_2$ on different substrates is compared in the supporting information (see Supplementary Information Figs. S2 and S3).

To further confirm the quality of the $MoS_2$ atomic layers prepared by our CVD method, Fig. 2a shows the morphology of an atomic $MoS_2$ layer covering on the TEM grid with a rolled-up edge, and Fig. 2b shows the edge area. Fig. 2c and 2d shows the two-layered and three layered $MoS_2$ samples. The interlayer spacing was measured to be ~6.6±0.2 Å. Fig. 2e and 2f are HRTEM of $MoS_2$ atomic layers. Circle in 2e indicates the Moiré patterns. The hexagonal structure could be clearly found in Fig. 2f. Fig. 2g and 2h are diffraction patterns, showing single-layered and double-layered areas. Fig. 2i, 2j and 2k shows elemental mappings. Fig. 2i is the original images and Fig. 2j and 2k are Mo and S elemental mappings, respectively. The EELS results are also shown in Fig. 2l and 2m. The EELS spectrum obtained from the location, indicated by the red dot in Fig. 2i, reveals the characteristic peaks of Mo at 35 eV (N-edge) and S at 165 eV (L-edge)[13]. The ratio of Mo and S is about 1:2, which is confirmed by the XPS data (see Supplementary Information Figs. S4).

The grain size of CVD-grown and liquid mechanical exfoliated $MoS_2$ (LE-$MoS_2$),[9] as a comparison, could be estimated by the dark-field (DF) TEM images shown in Figure 3. Fig. 3a shows a bright-field (BF) TEM image of a random area in the CVD $MoS_2$. Fig. 3b and 3c are corresponding diffraction pattern and false-color DF TEM image of area in Fig. 3a, suggesting a poly-crystalline $MoS_2$ with a grain size ranging from 10 nm ~ 30nm. Fig. 3b contains multi-group six-fold-symmetry spots, which is also seen in CVD graphene.[14] The false-color DF TEM image is taken using an objective aperture filter to cover three spots in the back focus plane, marked by the circle. The colors (red, green and blue) in the DF TEM image correspond to the ones of circles in Fig. 3b. Fig. 3d-3e are the BF TEM image, diffraction pattern and DF TEM image of LE-$MoS_2$, respectively. The individual six-fold-symmetry pattern suggests the grain size of is larger than 1 μm or more. This result is further confirmed by the comparison of random edges in CVD and

LE MoS$_2$, as shown in Fig. 3g and 3h. It can be found 4L and 3L in length of ~10nm, 2L MoS$_2$ in length of ~20nm in CVD MoS$_2$, and 4L in length of ~90nm in LE MoS$_2$.

Raman spectra on as-prepared MoS$_2$ atomic layers, as well as mechanically exfoliated thin flakes were collected for comparisons. As shown in Figure 4, Raman spectra were collected for single-layered and double-layered MoS$_2$ samples on SiO$_2$ substrate. Two typical Raman active modes could be found: $E^1_{2g}$ at 383 cm$^{-1}$ and $A_{1g}$ at 409 cm$^{-1}$ [15]. These modes of vibration have been investigated both theoretically and empirically in bulk MoS$_2$[16-18], $E^1_{2g}$ indicates planar vibration and $A_{1g}$ associates with the vibration of sulfides in the out-of-plane direction as illustrated in the inset of Fig. 4a. Some criterion could be used to roughly identify the thickness of the layers[15]: (1) Raman peak location and intensity of $E^1_{2g}$ and $A_{1g}$ (with same parameters like laser power, collecting time etc.). The peaks were found to be blue-shift for $E^1_{2g}$ and red-shift for $A_{1g}$ when the film becomes thinner, which would also result in a weaker signal. In Figs. 4a and 4b, their peaks from $E^1_{2g}$ and $A_{1g}$ located at 384.6 cm$^{-1}$, 405.1 cm$^{-1}$ and 384.6 cm$^{-1}$, 406.9 cm$^{-1}$, respectively, which corresponded to single-layered and double-layered MoS$_2$ samples. The spectra in blue are recorded from mechanical exfoliated MoS$_2$ with a corresponding numbers of layer[15]; (2) The peak spacing between $E^1_{2g}$ and $A_{1g}$. In our case, they were 20.6 cm$^{-1}$ for single-layered and 22.3 cm$^{-1}$ for double-layer samples; (3) The intensity ratio between the characteristic peaks from MoS$_2$ and the substrate. For our samples, $E^1_{2g}$/Si were ~ 0.05 and 0.09, again corresponding to single-layered and double-layered MoS$_2$ samples[15]. The Raman intensity ratios for $E^1_{2g}$ and $A^1_g$ are different for the CVD MoS$_2$ and exfoliated MoS$_2$. It is because the planar vibration ($E^1_{2g}$) is subject to the nano-scale and random-distributed grains in CVD MoS$_2$ (Fig. 3c), therefore showing a lower relative intensity compared to mechanical exfoliated MoS$_2$. It is supported by further studies on the DF TEM image of exfoliated MoS$_2$ flakes. Their grain size is much larger, typically at the order of microns or more (Fig. 3f). Raman mapping was taken from the dashed area (35 μm×45 μm) shown in Fig. 4c, which is a typical edge area of a large size atomic MoS$_2$ layer prepared by our CVD method. Fig. 4d and 4e represent the intensity mapping ($E^1_{2g}$) and intensity ratio mapping ($E^1_{2g}$/Si). There were total 576 (24×29) Raman spectra collected from this area. Both mappings show a similar landscape. Intensity ratio mapping provides a more accurate characterization and better resolution for the atomic layer samples with different thicknesses. The thin area was shown in light blue and thick area in red. Raman spectra strongly suggest good quality, uniform coverage of $MoS_2$ atomic layers (from single layer to a few layers) on $SiO_2$ substrate.

Field effect transistor (FET) devices were made by photolithography process to determine the electric transport properties of CVD-prepared $MoS_2$. We use photoresists S1813 and LOR5B to make electrodes patterns with under-cut structures by mask aligner (SUSS Mask Aligner MJB4) and then develop with MF319. Ti/Au Electrodes (5 nm/30 nm) are deposited by e-beam evaporator. The evaporating rate was well controlled about 1 Å/s. The photoresist could be removed by acetone and PG-REMOVER. The electrical measurements were carried out using two Keithley 2400 source meters connected with a CTI Cryodyne Refrigeration System to provide a temperature ranging from 15K to 450K and a vacuum down to $7 \times 10^{-6}$ Torr. Their electrical transport properties are shown in Figure 5. Fig. 5a is a typical device with an electrode spacing ~ 9 μm and the length of the electrodes is ~100 μm. Fig. 5b is a typical I-V curve of $MoS_2$ device with a resistance of ~130 KΩ. For most of the devices, their source current versus bias voltage is linear ranging from 1mV to 1V, suggesting ohmic contacts with our Ti/Au electrodes. The resistivity of our $MoS_2$ samples are from ~ $1.46 \times 10^4$ Ω/□ to $2.84 \times 10^4$ Ω/□, about two orders of magnitude higher than the CVD-prepared graphene (125Ω/□).[19] Temperature dependence measurement indicates that $MoS_2$'s resistance increases at low temperatures, as shown in Fig. 5c. The typical mobilities measured are ranging from 0.004 to 0.04 $cm^2V^{-1}s^{-1}$ at room temperature, one to two orders of magnitude less than the mechanical exfoliated $MoS_2$ samples (0.1 ~10 $cm^2V^{-1}s^{-1}$).[1] The mobility of $MoS_2$ at low-field field effect is estimated by $\mu = [dI_{ds}/dV_{bg}] \times [(L/(WC_iV_{ds}))]$. Here $L$ is the channel length ~9 μm, $W$ is the channel width from 17 μm to 80 μm for various devices. $C_i$ ~ $1.3 \times 10^{-4}$ F $m^{-2}$ is capacitance between the channel and the back-gate per unit area. We believe the low mobilities originate from the planar defect - the nano-scale and random-distributed CVD $MoS_2$ grains, as shown in the DF TEM image in Fig. 3c. Electron hopping among grains would significantly decrease the mobilities in $MoS_2$.[22, 23] In addition, other defects including cationic vacancies, dislocation and adsorption-induced doping effect in the $MoS_2$ are also possible reasons for the low mobilities, which are always observed in

CVD-prepared two dimensional materials like graphene.[14] The mobility could be significantly improved by annealing the as-prepared samples,[8, 20] using local top-gate with high-κ dielectric,[8, 21] and optimizing the growth conditions. Different from the naturally grown $MoS_2$ crystal that is n-type semiconductor, we observed that our CVD-prepared $MoS_2$ is an intrinsic p-type semiconductor at room temperature, as shown in Fig. 5d. Further work would be required to clarify such differences.

The reaction mechanism for synthesizing $MoS_2$ atomic layers could be simply understood as a direct elemental chemical reaction. In our experiments, the earlier reported precursors used in synthesizing $MoS_2$ nanostructures[22-27] were not selected, since it's very difficult to obtain large area uniform film from those precursors. Metal substrates have also been considered in experiments. In fact, the reactions between S and metals at relevant reaction temperatures make Au almost the only suitable metal substrate. The resulted $MoS_2$ atomic layers grown on such substrate display many interesting tent-like microstructures (see Supplementary Information Figs. S5 and S6). These suspended, perhaps pre-stressed atomic layers could have some unique properties and also help us learn more about mechanical properties of such atomic-layered $MoS_2$ samples.

In summary, we have shown here a direct preparation of monolayer and few-layered $MoS_2$ on $SiO_2$ substrates using a pre-deposition of Mo film followed by CVD method. The size and thickness of atomic $MoS_2$ layer depend on the size of the substrate and the thickness of the pre-deposited Mo, which are easily scalable and controllable, making it possible to meet the demands from different applications. Characterization such as HRTEM and Raman indicate the as-prepared $MoS_2$ are of good quality and crystallinity, and ranges typically from mono-layer to a few layers. Our new large area synthesis method has thus revealed new possibility to prepare large area good quality $MoS_2$ atomic layer materials, increasing the number of possible candidates to be engineered into 2D structures in the direction provided by the advent of graphene and its applications.

**Acknowledgements**

J. L. acknowledges the support by the Welch Foundation grant C-1716, the Air Force Research Laboratory grant AFRL FA8650-07-2-5061 and the NSF grant CMMI 0928297. P.M.A. acknowledges support from Rice University startup funds, and P. M. A and Z. L. acknowledge funding support from the Office of Naval Research (ONR) through the MURI program on graphene. The authors would like to acknowledge Mr. Yusuke Nakamura for his help on CVD growth and Mr. Jiangnan Zhang for his help on Mo film thickness measurements.

**Supporting Information Available:** Description of CVD setup, additional Raman and XPS characterization results and $MoS_2$ atomic layer growth on metal substrates. This material is available free of charge via the Internet at http://pubs.acs.org.

# Figure Legends

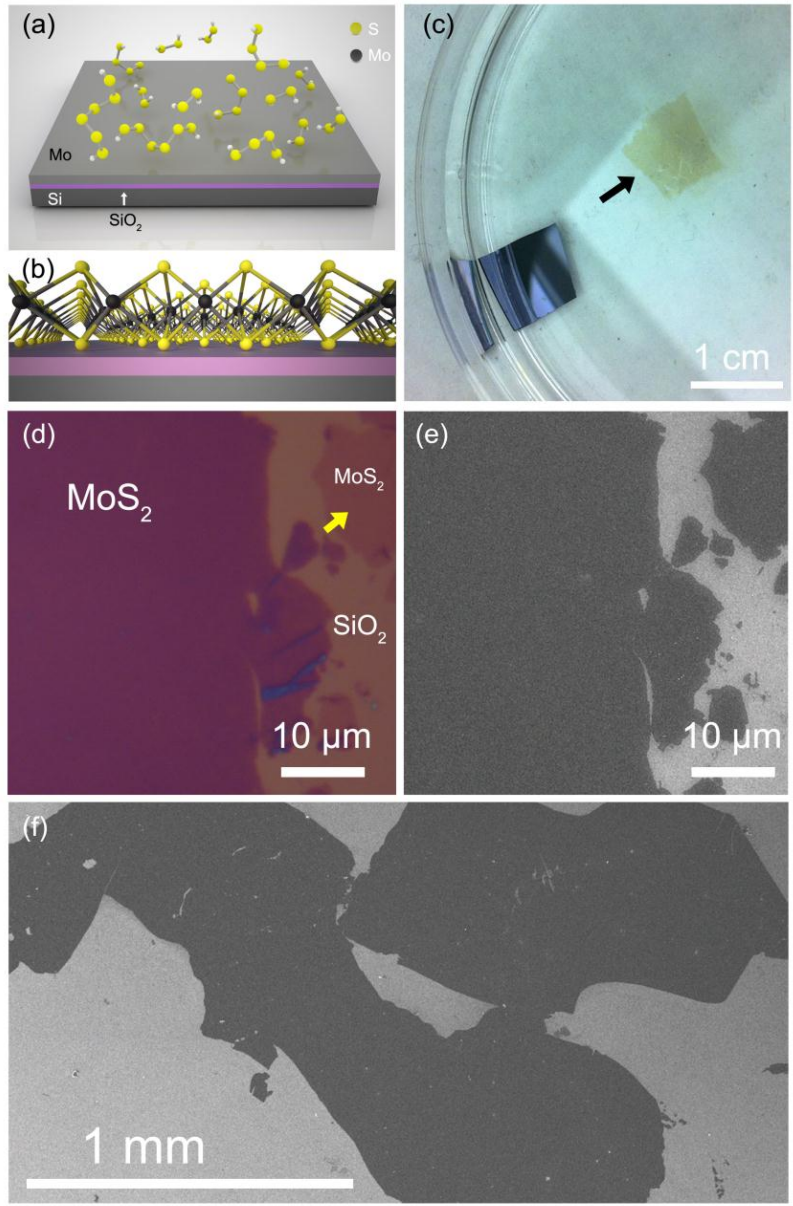

**Figure 1 Illustrations and morphologies of atomic layered MoS$_2$. a**, Introducing sulfur on Mo thin film that was pre-deposited on SiO$_2$ substrate; **b**, MoS$_2$ films that are directly grown on the SiO$_2$ substrate. The atoms in back and yellow represent Mo and S, respectively; **c**, SiO$_2$/Si substrate (left) and peeled off few layer MoS$_2$ (right, indicated by the arrow) floating on KOH solution; **d**, Optical image of one local section with MoS$_2$ on SiO$_2$/Si substrate. Most of areas in purple are few-layered MoS$_2$. The area in light purple is 1-2 layered MoS$_2$ marked by a yellow arrow; **e**, Corresponding SEM image. These images show a large size, uniform and continuous MoS$_2$ atomic layer. **f**, SEM image of large area MoS$_2$.

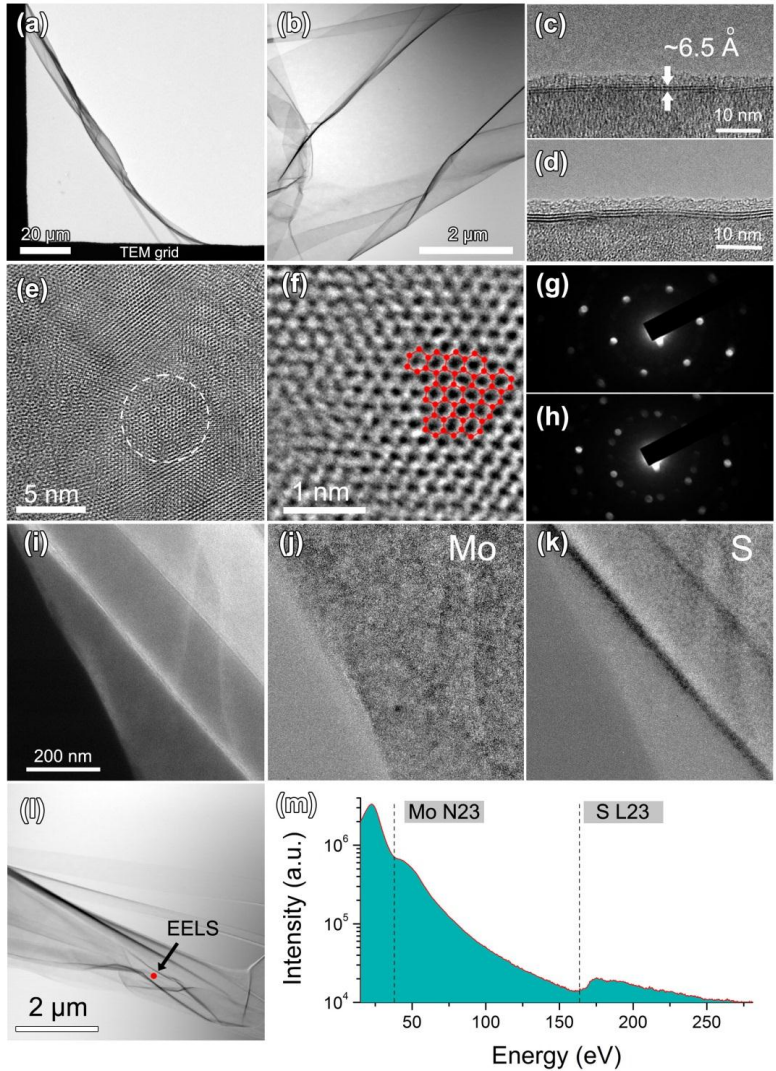

**Figure 2 TEM characterizations and chemical elemental analysis of CVD-grown MoS$_2$. a**, One atomic MoS$_2$ layer covers on the TEM grid; **b**, Edge area of the atomic MoS$_2$ layer in **a**; **c-d**, Two and three layers of MoS$_2$. The distance between two layers is about 6.5Å; **e**, HRTEM images. The area marked by a circle in e shows the Moiré patterns; **f**, Atomic image of the MoS$_2$ layer shows a typical hexagonal structure. **g-h**, Diffraction patterns of the atomic layers; **i-k**, Original phase contrast image and corresponding molybdenum and sulfur elemental mappings, indicating the uniform distribution of Mo and S elements in the atomic layer; **l-m**, EELS shows the Mo edge and S edge at ~35eV and ~165eV, respectively. The red dot indicates the area where EELS data was collected.

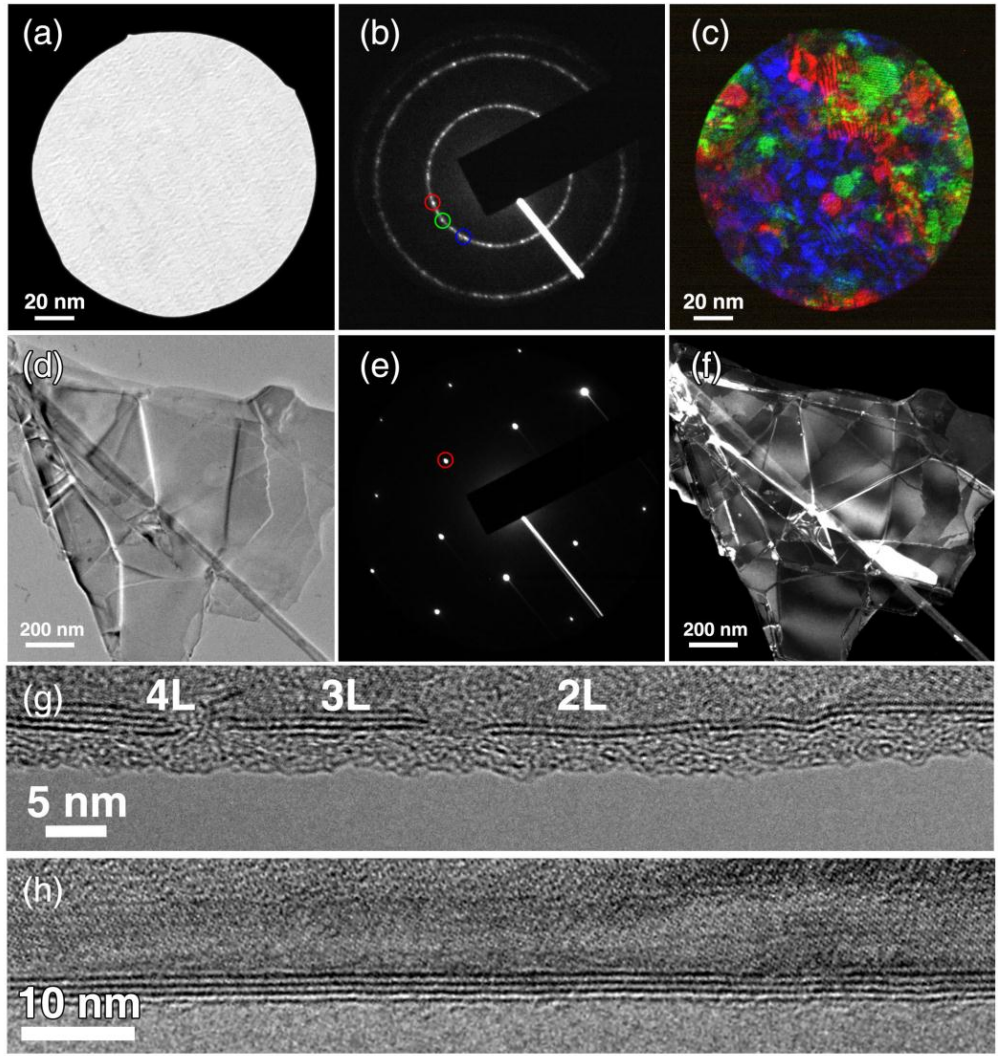

**Figure 3 Comparison of grain size in CVD-grown and naturally formed MoS$_2$. a,** Random area of CVD-grown MoS2 appear uniform in bright-field TEM images, **b,** Diffraction pattern taken from of area in **a** show the MoS2 is polycrystalline, **c,** a dark-field image corresponding to **a** with false color, **d,** Bright-field liquid exfoliated MoS$_2$ flake, **e,** Diffraction pattern taken from a region in **d** showing a single crystal MoS$_2$, **e,** A corresponding dark-field image, **g** and **h,** Typical edges of CVD MoS$_2$ and liquid exfoliated MoS$_2$.

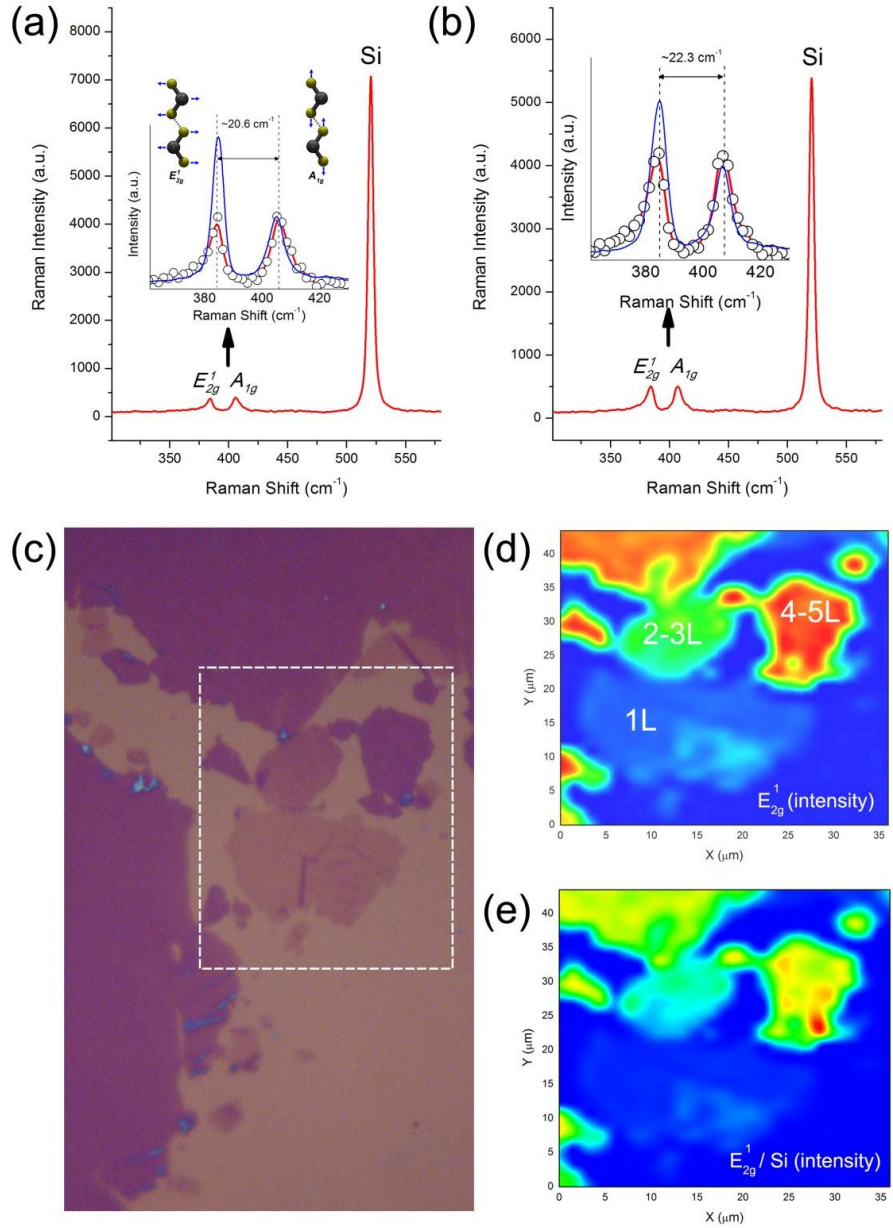

**Figure 4 Raman signatures of as-prepared CVD MoS$_2$ atomic layers. a-b**, Raman spectra of single-layered and double-layered MoS$_2$. The thickness of MoS$_2$ layers can be estimated by evaluating their relative intensity to Si, or the spacing between two vibrating modes (E$^1_{2g}$ and A$_{1g}$), as shown in the inset. Spectra in blue in the inset are from mechanical exfoliated MoS$_2$ (single-layered MoS$_2$ in **a** and double-layered in **b**; **c,** A typical landscape of MoS$_2$ atomic layers on SiO$_2$ substrate. The dotted area is mapped in **d**) (intensity of E$^1_{2g}$ peak) and **e** (E$^1_{2g}$ (intensity)/Si(intensity)), indicating the number of layers.

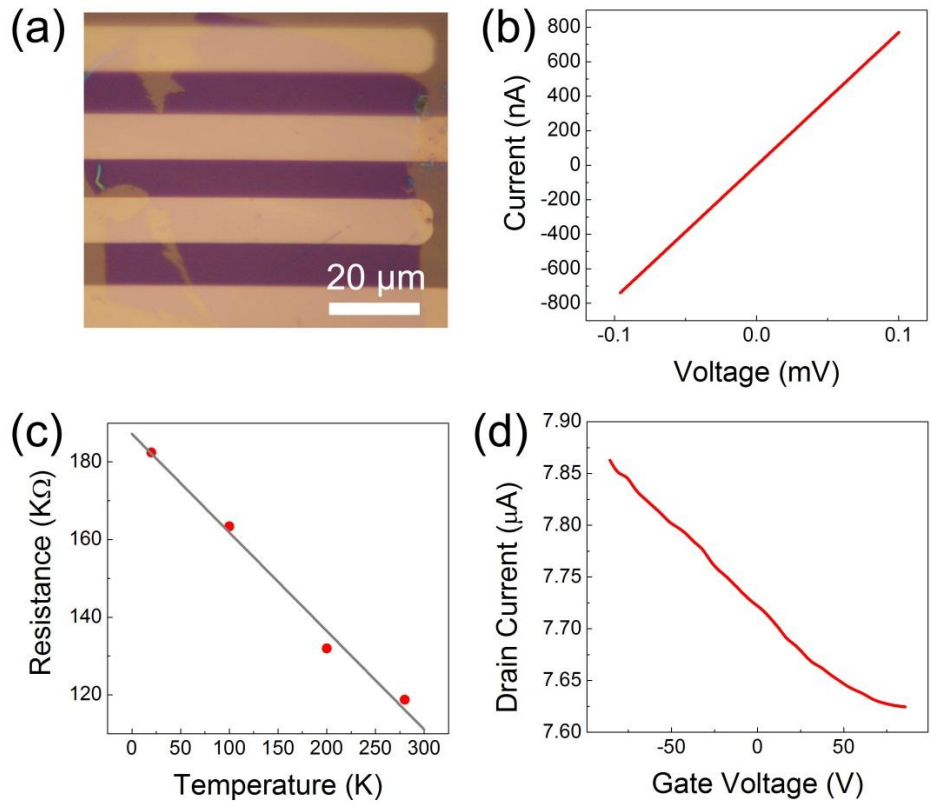

**Figure 5 Characterizations of MoS$_2$ devices. a**, Optical image of a typical MoS$_2$ device; **b**, $I_{ds}$-$V_{ds}$ curve acquired without a gate voltage; **c**, Temperature dependence of the resistance from 300K to 20K; **d**, Gate voltage versus drain current shows an intrinsic p-type MoS$_2$.

# Supporting Information for

# Large Area Vapor Phase Growth and Characterization of MoS$_2$ Atomic Layers on SiO$_2$ Substrate

Yongjie Zhan[1,#], Zheng Liu[1,#], Sina Najmaei[1], Pulickel M. Ajayan[1]* & Jun Lou[1]*

1. Department of Mechanical Engineering & Materials Science, Rice University, Houston, Texas 77005, US

# 1. Optical and SEM images of CVD MoS$_2$

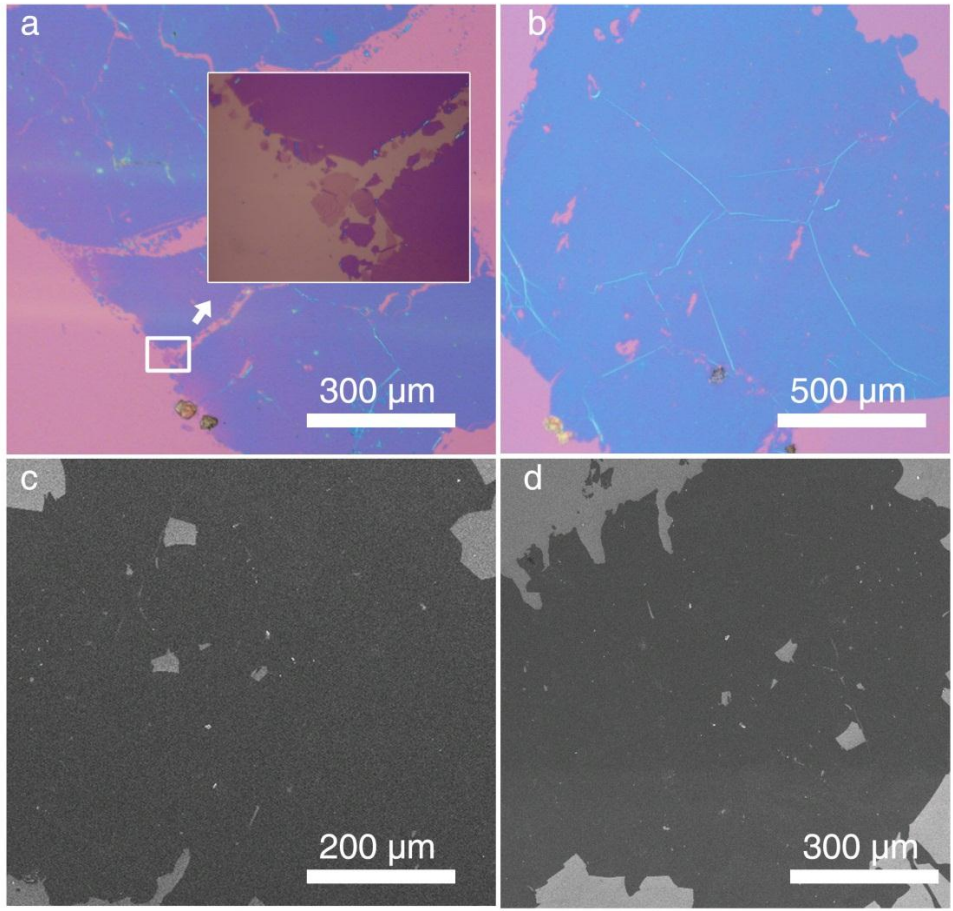

Figure S1. (a) and (b), Optical images of CVD-grown MoS2. Inset in a: An zoom-out area marked by a white arrow. (c) and (d), SEM images of MoS$_2$. The MoS$_2$ size can be easily scalable to the order of millimeters.

## 2. Schematic of the chemical vapor deposition (CVD) system.

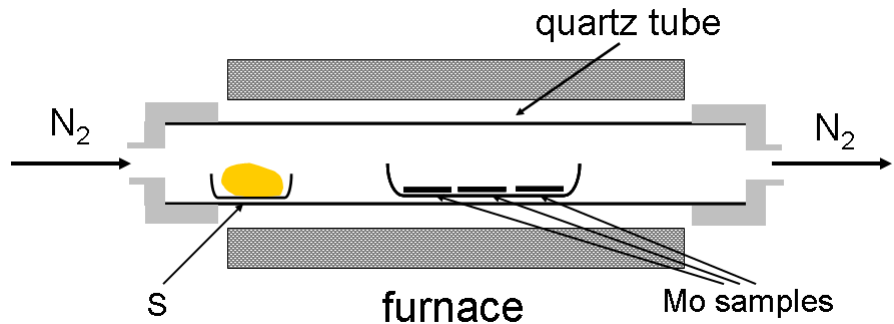

Figure S2. The CVD system to prepare $MoS_2$ samples

Mo thin films deposited on $SiO_2$ substrates were placed in a ceramic boat and then loaded into the center of a tube furnace. Pure sulfur in another boat was placed at the upwind low temperature zone in the same quartz tube. During the reaction, the temperature surrounding sulfur was kept to be slightly above its melting point ~113$^o$C.

The quartz tube was first kept in a flowing protective atmosphere of high purity $N_2$, the flow rate of was ~ 150-200 sccm (standard cubic centimeters per minute). After 15 minutes of $N_2$ purging, the furnace temperature was gradually increased from room temperature to 500 $^o$C in 30 minutes. Then the temperature was increased from 500 $^o$C to 750 $^o$C in 90minutes and was kept at 750 $^o$C for 10 minutes before cooled down to room temperature in 120 minutes. Figure S2 shows an illustration of the reaction condition of this CVD process.

## 3. Raman spectra of CVD MoS$_2$ grown on various substrates

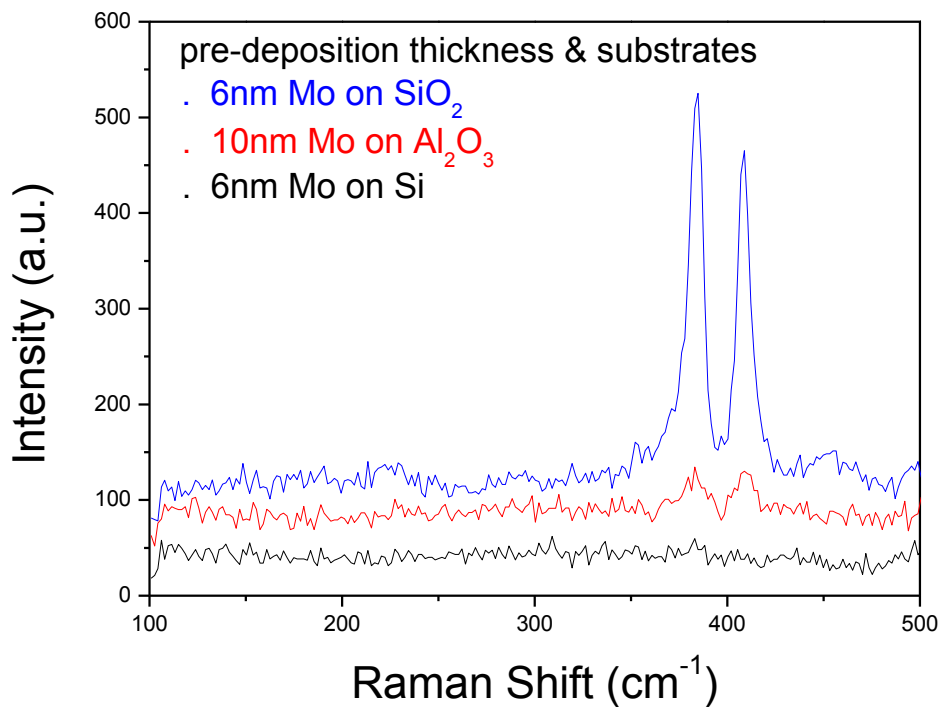

Figure S3. Raman spectra of MoS$_2$ samples grown on different substrates.

Raman spectroscopy is used to identify the quality of CVD MoS$_2$ films grown on 3 different substrates with a 514.5 cm$^{-1}$ laser. The peaks locate ~385 cm$^{-1}$ correspond the $E^1_{2g}$ vibration mode of MoS$_2$, and peaks at ~408 cm$^{-1}$ correspond to the $A_{1g}$ mode.[1] It can be found that thin MoS$_2$ samples can be grown on various substrates including SiO$_2$, Au, Si et al. The Raman signal is weak for MoS$_2$ on Si.

## 4. XPS spectra of CVD MoS$_2$

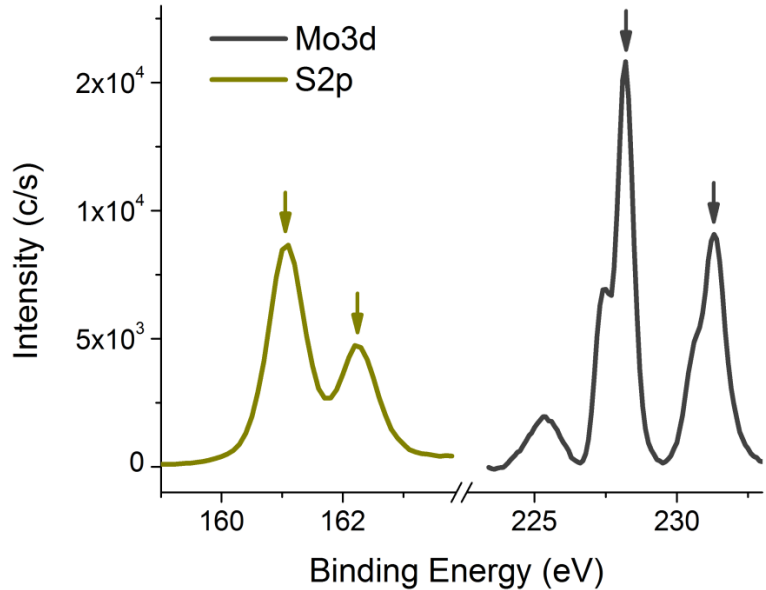

Figure S4. XPS spectra of the MoS$_2$ thin film showing the typical Mo and S peaks from MoS$_2$.

The XPS spectra of the as-grown MoS$_2$ film for the Mo and S edges are shown in Figure S2. Sulfur is in brown color. It shows 2p1/2 and 2p3/2 core levels at 162.3 eV and 161.2 eV, respectively, marked by the arrows, close to the previous reports (2p1/2: 164.1 eV,[2], 2p3/2: 161.5 eV ~ 163.4 eV[2-4]). The spectrum Molybdenum is in black. The Mo 3d3/2 and 3d5/2 peaks are around ~231.3 eV and ~228.2 eV, indicated by the black arrows, which is almost identify to the bulk MoS$_2$ samples (3d3/2: 232.3 eV ~ 233.3 eV, 3d5/2: 228.8 eV ~ 230.1 eV)[2,5,6] The calculated atomic concentration of S and Mo are 68.49% and 31.51%, with a ratio close to 2:1.

## 5. Syntheses of MoS₂ films on Au substrate and Raman Sepctrum

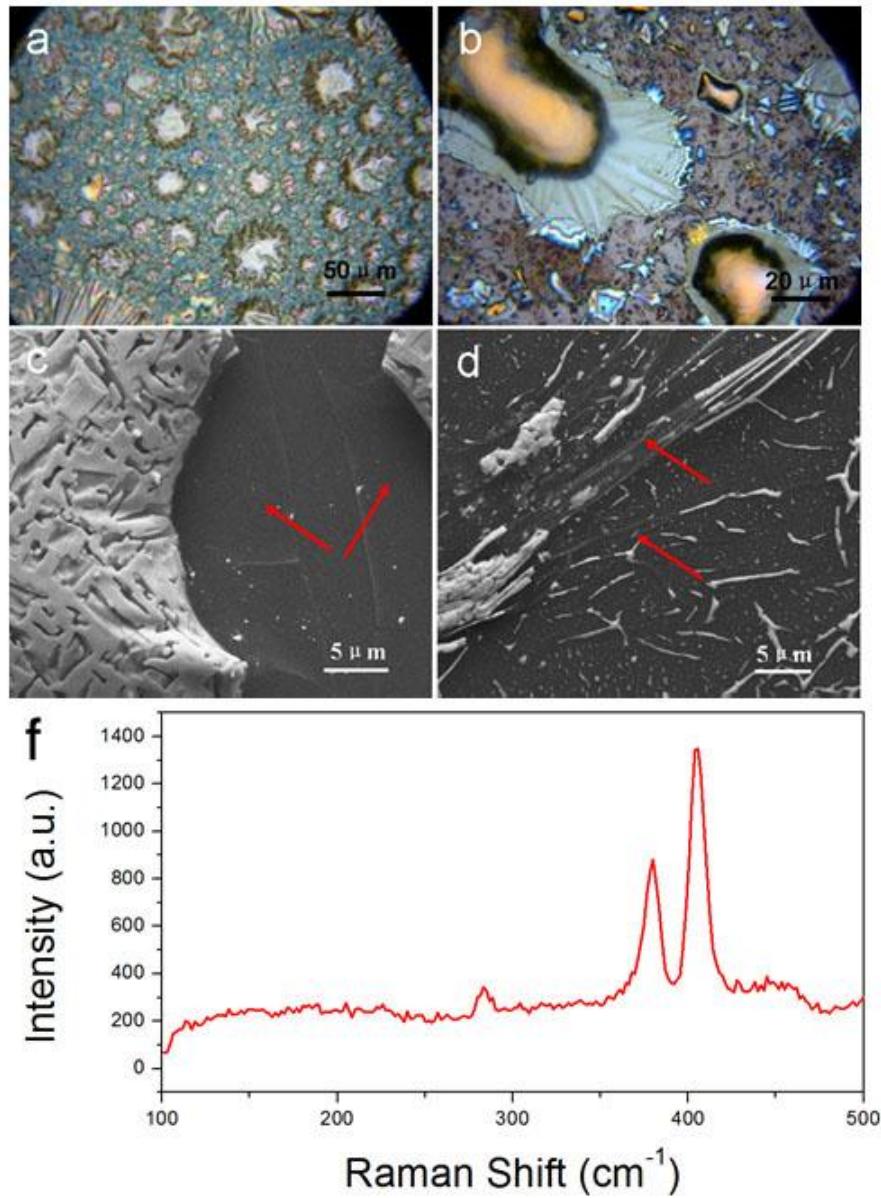

Figure S5. (a) and (b). Optical images of CVD MoS$_2$ films on Au substrates. The yellow parts are Au particles. (c) and (d) SEM image of MoS$_2$ films marked by the red arrows. (f) Raman spectrum of MoS$_2$ on Au films.

Au is an inert metal and does not react with sulfur in during synthesis of MoS$_2$. The thicknesses of gold films are proved to be a key factor in our experiments. Thickness below 100 nm was not thick enough and would shrink into isolated micro-balls on silicon substrate after the annealing process during synthesis. Au films with a thickness of ~350nm are finally determined.

Figure S5 shows optical, SEM images and Raman spectrum of typical MoS$_2$ samples grown on Au substrate with a thickness of 350nm. The Mo thickness is ~ 3 nm. After high temperature annealing, Au substrate shrank into particles (Figure S5b). The MoS$_2$ films can be found on most of areas marked by the red arrows (Figure S5c and S5d). Raman spectra show the $E^1_{2g}$ and $A^1_g$ mode of MoS$_2$. As shown in SEM images, red arrows reveal more details of these films surrounding Au islands and on Au substrate. Also, the suspended MoS$_2$ film in Figure S5d seems like very thin as they are transparent. Thanks to the highly conductive Au substrate, the MoS$_2$ films are much clearer under SEM than those grown on SiO$_2$ substrate.

## 6. Formation of suspended MoS$_2$ film.

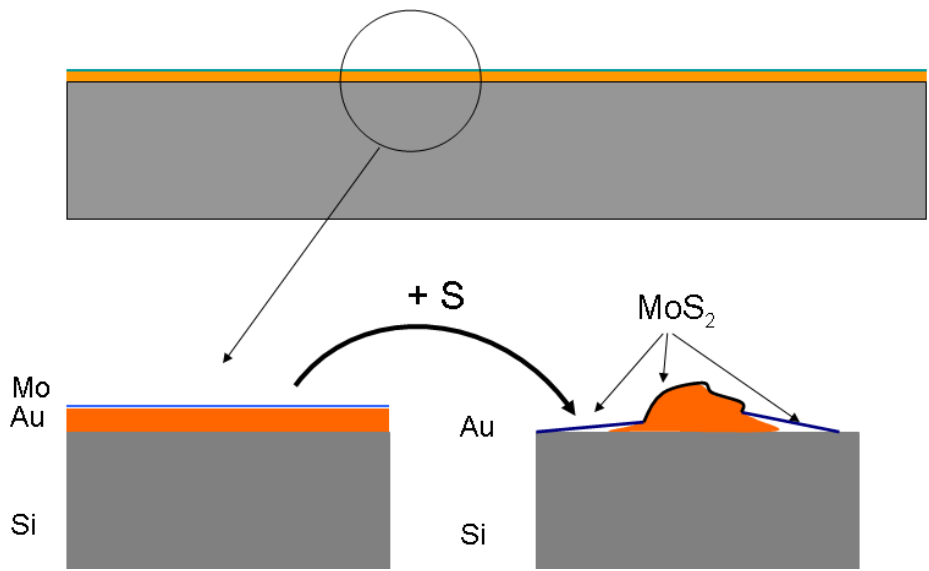

Figure S6. Illustrations of the formation of suspended MoS$_2$ film.

The Au and Mo layers are deposited by sputtering and E-beam evaporator, respectively. The MoS$_2$ film is formed before the Au film shrinks into particles. During the annealing process (750 $^o$C for 10min), the MoS$_2$ films are deformed when the gold film shrink into particles, forming a suspended MoS$_2$ film (Fig. S5d).